# Humboldt: Metadata-Driven Extensible Data Discovery

Alex Bäuerle*  
AxiomBio

Çağatay Demiralp*  
Amazon & MIT

Michael Stonebraker  
MIT

## ABSTRACT

Data discovery is crucial for data management and analysis and can benefit from better utilization of metadata. For example, users may want to search data using queries like "find the tables created by Alex and endorsed by Mike that contain sales numbers." They may also want to see how the data they view relates to other data, its lineage, or the quality and compliance of its upstream datasets, all metadata. Yet, effectively surfacing metadata through interactive user interfaces (UIs) to augment data discovery poses challenges. Constantly revamping UIs with each update to metadata sources (or providers) consumes significant development resources and lacks scalability and extensibility.

In response, we introduce Humboldt, a new framework enabling interactive data systems to effectively leverage metadata for data discovery and rapidly evolve their UIs to support metadata changes. Humboldt decouples metadata sources from the implementation of data discovery UIs that support search and dataset visualization using metadata fields. It automatically generates interactive data discovery interfaces from declarative specifications, avoiding costly metadata-specific (re)implementations.

To evaluate Humboldt, we implement it in a commercial SaaS application for interactive business data analysis. We demonstrate its expressiveness by automatically generating a new feature-rich data discovery interface in the application using a few lines of Humboldt specification. This new interface offers several types of discovery views based on the characteristics of the available data and metadata. It also supports composable query-based search and automatically enables queries such as "*type*: table *owned by*: 'Alex' *badged*: endorsed *badged by*: 'Mike' & 'sales'" where query parameters are compiled from the specification. We also evaluate how well the discovery interface benefits end users through a preliminary study conducted with sales engineers. Results show that the Humboldt-generated interface assists users in effectively utilizing metadata for discovery and search, with varying ease in using its components. The results also suggest organizations and users have varying data discovery needs and preferences, validating Humboldt's design goals of expressivity, composability, and configurability.

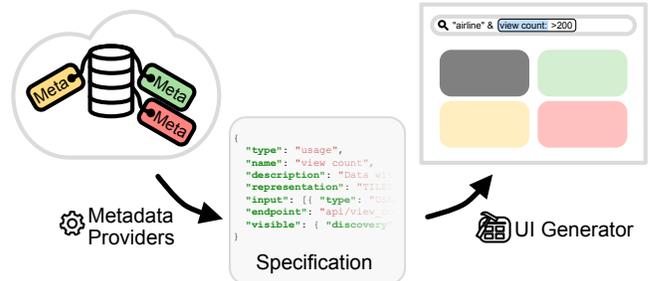

**Figure 1:** The Humboldt[1] framework enables interactive data systems to effectively leverage metadata for data discovery and rapidly evolve their user interfaces through declarative specification to support metadata changes.

## 1 INTRODUCTION

The development of easy-to-use data analysis tools [14, 20, 31] and the low cost of storing data in public cloud [1] provided broad audiences with the possibility of accessing and making use of large collections of data without writing code or relying on data analysts. On the other hand, the extended usability also led to data stores in enterprises with a much less modeled state and a rapid expansion of the data landscape with the proliferation of derived artifacts (e.g., tables, visualizations, dashboards) [30]. As a result, finding the right data for the task (data discovery) has become even more challenging. Metadata such as data ownership, usage, certification, annotations, dictionaries, and relationships (primary-foreign key, lineage, downstream data flow, semantic similarity, etc.) can play an essential role in addressing the challenge by providing business and semantic context to constrain the search space [13]. For example, metadata can enable an employee who recently joined the marketing department to find the marketing attribution dashboard endorsed by the manager and frequently viewed by the team members. The employee can further check the lineage of the data underlying the found dashboard to get a quick sense of what tables to trust.

Some commercial data analysis tools (e.g., [28, 33]) offer APIs to query content using metadata. However, these services are inaccessible to users who don't code and, crucially, separate metadata from the interactive data analysis context. Discovery user interfaces (UIs) should surface metadata effectively to be useful for broader audiences. For example, the new employee above should be able to filter the available dashboards based on their endorsement tags and endorsers, e.g., by selecting the condition from a dropdown menu or running a keyword search. Yet, existing interactive data systems have limited support for metadata-driven data discovery. The UIs for data discovery using metadata in these systems are typically implemented using hardcoded views and interactions, where any update to the metadata sources requires expensive and error-prone code changes. Therefore, UIs that enable interaction with new metadata and derived representations are often implemented long after metadata becomes available. Even then, they are difficult to customize and extend to support the varying data discovery needs of users and domains.

---

[1] See a demo video of Humboldt at https://tinyurl.com/4csh4axj

In response, we introduce Humboldt (Figure 1) to dynamically generate data discovery UIs from declarative specifications, enabling easy integration and use of different types of metadata in interactive data systems without going through costly software upgrade cycles. In Humboldt, enabling new metadata for data discovery and search is just a matter of adding a few lines of specification instead of changing the UI implementation. Consider an admin who wants to make the relationship metadata based on a machine learning model on table similarity available for data discovery in the UI. Adding the model as a new metadata provider[2] in Humboldt's specification would suffice to enable such support with the relevant views and visualizations generated automatically.

To find the right data, users need contextual views (e.g., *"which dashboards are my teammates working on?"*), exploration tools (e.g., *"show me data that is joinable to what I'm looking at"*), and filters (e.g., *"show me only analyses from a specific user"*). Humboldt's specification supports these tasks through three main data discovery features: overviews, exploration, and search. From the specification, we generate various views that serve as entry points to data analysis. When interacting with data, Humboldt can surface related data artifacts, facilitating further exploration. Humboldt also builds a query language for complex, metadata-based search and filtering. Serving as a layer of abstraction on top of metadata providers, Humboldt makes it easy to add, change, or remove metadata providers without changing any UI code.

Humboldt is designed as an interface between existing metadata providers and a data discovery UI. Adding new metadata providers or changing their behavior does not require changes in the UI code. At the same time, Humboldt's representation of metadata providers enables an automatic generation of UI elements for data discovery. In summary, our contributions are as follows :

- We introduce Humboldt, a specification-based system framework for generating a data discovery UI from different metadata providers. We implement this new framework in Sigma Workbook [14], a commercial SaaS application for business data analysis.

- We demonstrate the expressivity of the Humboldt framework by automatically generating an interactive UI in Sigma Workbook, supporting multiple search paradigms, different types of views (Figure 6), composable queries, and ranking algorithms for metadata-driven data discovery.

- We evaluate the new data discovery UI in Sigma Workbook generated by Humboldt in a user study with six users at Sigma Computing. The generated UI effectively integrates metadata for improved data discovery and search. Different teams and users have different data discovery needs and preferences, further motivating the Humboldt design. Humboldt enables data systems to tailor their data discovery UIs to users' needs and preferences.

---

[2]Metadata provider is a metadata source, typically an API endpoint.

## 2 RELATED WORK

Our work relates to two categories of prior data discovery research. Research in the first category concerns techniques for automatically extracting and computing the properties of and connections between datasets (e.g., descriptive, structural, or provenance metadata) for augmenting data discovery. The second category of earlier research focuses on interactive interfaces for data discovery. We summarize both lines of related work below.

**Data Discovery Techniques.** Earlier work computes how tables in a database relate to each other using various similarity measures. Most similarity computations operate on descriptors or signatures of table columns (e.g., MinHash sketches, TF-IDF, headers, embeddings, sample values, etc.). These descriptors and signatures determine the type of relationship computed. Types of similarities captured by earlier work include, e.g., value overlap [5, 6], distribution [6], examples [15, 27], and semantic similarities [7, 10, 12, 17]. Others use table-wide measures such as schema [11] and entity complement [11, 35] to identify joinable and unionable tables for a given query table. Earlier research also uses column similarity to build lifted representations such as graphs to support efficient dataset queries [6, 17] and navigation [21].

The similarity or relatedness of one dataset to another is often task and domain-dependent. Prior work such as D3L [3] and Voyager [4] combines ensembles of similarity measures for data discovery. This ensemble approach improves over, for example, Aurum [6] on table union search.

Modeling and computing relevance are necessary for data discovery but orthogonal to Humboldt. The earlier work exemplified above focuses on approaches for extracting and computing relevance (a form of metadata) for data discovery. It leaves how to surface it to users largely unexplored, which Humboldt aims to address. We designed Humboldt to facilitate extensible and evolvable data discovery UIs in interactive systems that can easily integrate and compose different types of metadata for augmenting dataset search [2].

**Data Discovery Interfaces.** Data discovery is a form of search [9], and there is extensive literature on improving the user experience for search [16]. Exploratory search [18, 34] is essential for data discovery where users may not know what they are looking for. Prior work uses faceted browsers [36] and dynamic queries [29] to enable users incrementally to build partial specifications of search queries through direct manipulation.

Research on the Semantic Web has explored ways to create user-friendly interfaces for surfacing data specified in Resource Description Format (RDF) [25, 26]. While our approach also generates UI from a declarative specification, it focuses on integrating metadata sources for improving data discovery interfaces in the interactive enterprise data systems.

Recent work introduces data discovery systems focusing on UI-based visual interaction, bringing the two lines of prior research above closer together. Kyrix-J [32] automatically generates visual transitions (jumps) between prespecified multiscale visualizations of database tables, where a connection provider supplies the connections (similarities) between tables. Auctus [5] surfaces data profiles and relationships in an interface similar to web search interfaces for a user-friendly data discovery experience. With Ronin [23], users can start from a search query and explore related data in a hierarchical organization.

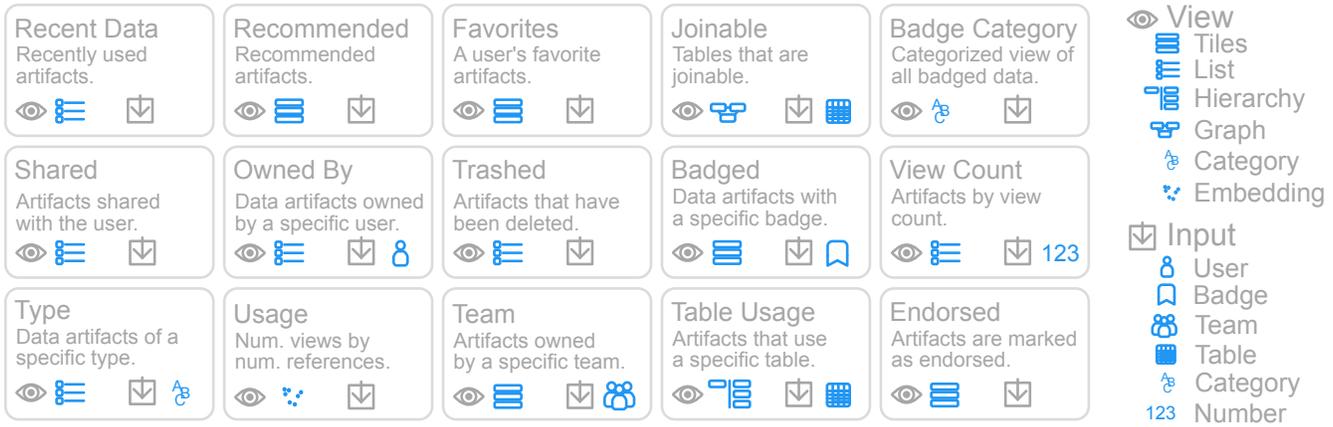

Figure 2: Examples of Humboldt metadata providers. Views generated for a metadata provider in the data discovery UI depend on the provider's specification, which may describe various aspects of the provided metadata, such as its source, data format, visual representation, visibility, field weights, and ranking.

However, earlier work on search interfaces and interactive data discovery systems provides limited means to configure and customize data discovery UIs, typically hardcoding the data search support based on descriptive, structural, or provenance metadata. In contrast, Humboldt enables data discovery that is adaptive to end-user needs and different considerations of what is relevant, allowing their quick, composable integration in the UI.

## 3 FRAMEWORK DESIGN

In the following, we describe the design process of Humboldt. Since two of us worked at Sigma Computing, a business intelligence (BI) company, we anchored our need-finding efforts around easy-to-use (or self-service) business data analysis tools, a large class of interactive data systems with significant practical impact. To inform our framework, we first conducted formative interviews to assess the needs of users. Subsequently, we derived design goals for a data discovery framework based on these insights.

### 3.1 Formative Interviews

To understand current BI tools' limitations concerning data discovery, we conducted seven semi-structured interviews at Sigma Computing. We recruited participants through internal communication channels such as email and Slack. Our participants worked in different roles, e.g., design, sales, and engineering. While some of our participants reflected more on their own experience with data discovery, many relayed customer feedback and, as such, revealed day-to-day problems with discovering relevant data in the field. Interviews were conducted via video calls and lasted for 30 minutes. We first asked our interviewees some general questions about their data discovery needs. Then, we inquired about the pain points they experienced when searching for data to answer their questions. Finally, we gave participants time to elaborate on potential solutions to their problems. From these interviews, we identified the following general themes.

**The need for data discovery.** The advent of modern interactive direct manipulation-based (i.e., no-code) BI tools made data analysis available for many business users who needed programming knowledge. However, the success of the analysis still depends on finding the right data. This is especially important when users have a large number of tables in their data stores, which can be *"up to millions."* (P3)

Often, users' first experience with a data analysis tool determines whether they will adopt it in their workflow as *"onboarding weeks makes users love [or hate] a product."* (P2) However, currently, a large portion of the initial onboarding process for BI tools is spent on searching for the right data to use so that the initial *"onboarding call for finding data takes about 30 min, which happens again for a user after the call is over."* (P2) In conclusion, our interviewees note that *"no-SQL analysis is great, but data needs to be found first"* (P6), underscoring the importance of data discovery.

**Effective points of entry for discovery.** Users *"don't have the motivation to browse hundreds of tables."* (P7) Hence, to start their analysis, users have to be able to obtain faceted overviews as windows into the data that is available to them. BI tools often collect data that would enable such faceting, e.g., *"frequency of use, recency, ownership, lineage, annotations, or other information"* (P1). However, providing various views as metadata-based windows into the data store is tedious and costly, as *"many still have a separate data modeling team, but it is really hard for them to keep up with the growing number of artifacts"* (P5). Furthermore, different users, or teams often have various needs that cannot be satisfied with a single, static UI.

**Linking information.** Understanding *"where [a table] belongs in the warehouse and what role it serves is very hard"* (P6). At the same time, this information can help users select *"the data set that I should use."* (P3) This is where linking information between different metadata providers comes into play. Without such links, *"even if you have a set of curated tables, you don't see the dashboard solving 90% of your problems"* (P5). Additionally, *"a maintainer of data may not know what has been done with it downstream"* (P4), which asks for *"more statistics about how data is used and how it interacts*

*with other data."* (P6) The need for such connections underlines the importance of providing various exploration paths based on the available metadata.

**Intelligent search and filters.** Finally, we found that *"even if you know what data is there, you don't always know where to find it."* (P5) Thus, a data discovery system needs to provide methods to search for data. However, users might be aware of different metadata information about the data they seek, such as ownership, usage metrics, names, etc. As such, *"a normal search bar is not enough for more complex queries"* (P6), and *"more detailed search to filter the search space"* (P6) is needed. Similarly, when surfacing data, there are often multiple similar data artifacts, to the point where *"of all the data that is very similar, I'm not sure which data set to use."* (P4) Hence, such metadata-based querying is essential not only for searching but also for filtering data.

## 3.2 Design Goals

Informed by our formative interviews and literature research, we formulated three primary design goals for Humboldt. Our goal is to integrate effectively and surface metadata for improved data discovery. We decouple metadata providers from the data discovery UI so that they can be quickly and independently updated, reused, and extended. To this end, our primary design goals for the Humboldt system framework were:

**Expressivity.** When providing an interface between metadata providers and the UI to present data, where views are generated automatically based on available data and metadata, there needs to be a specification of how data can be fetched and what response to expect. The different components of a data discovery UI, i.e., discovery, exploration, and search, can only react to changes in the underlying data representation if data is specified in a predefined format. For example, one can generate various views for data discovery, each of which might visualize data differently to provide optimal representations based on the metadata used as a window into the available data. This specification can also inform a query language that a search interface utilizes. Furthermore, new UI elements can be loaded when input values become available based on selected data artifacts.

**Composability.** While a specification of data abstractions helps inform a data discovery UI, it is not necessarily sufficient for linking information between metadata providers. However, connections between data elements are integral for further exploration from a given data source and complex search queries. For example, inspecting data involves common questions such as what the derived artifacts of the data are or what other data I can use that is similar regarding a specific metadata attribute. Additionally, one might want to use metadata-based filters to display a subset of the data presented by a view. This requires a composition of filter queries with the data provided by a metadata provider for that view.

**Configurability.** The number of metadata types and sources useful for data discovery can be large. We expect this number to only increase with automated (e.g., AI-driven) active metadata extraction approaches. Also, metadata collected in enterprises and techniques for determining what is relevant and related are often domain and

```
{
  "type": "joinable",
  "name": "Name-Based",
  "description": "Informs about joinable
  tables by looking at column names.",
  "representation": "GRAPH",
  "input": [
    { "type": "TABLEID", "required": true }
  ],
  "endpoint": "api/name_joinability",
  "visible": {
    "discovery": true,
    "search": true,
  },
}
```

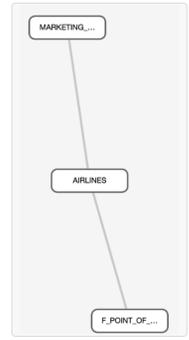

Figure 3: Left: An example of a metadata provider specification in Humboldt. Right: The resulting visualization in the data discovery UI. This metadata provider requires a table as input and returns a graph representation of joinability for the input table.

business-specific. A visual interface for data discovery must be adaptive and customizable to integrate and reap the benefits of different forms of metadata and data discovery techniques. Therefore, the third main design goal for Humboldt was the framework's configurability, including extensibility, to support a set of metadata providers that can be changed whenever new techniques become available. While this configurability helps developers who want to add or test new functionality, it can also be helpful for end users of the data discovery interface. Therefore, end users, particularly admins of data systems, should be able to configure the integration of the metadata providers most helpful for their analysis tasks.

## 4 HUMBOLDT SPECIFICATION

To formalize data representations in Humboldt, we use a specification based on which views are generated and which informs the interaction design of the data discovery system. Our motivation for using such a specification to inform data discovery follows many of the decision factors mentioned by Mernik et al. [19], as we aim to facilitate better automation, enable better product line architecture sharing, provide a generic representation of the underlying data structure, facilitate interaction, and inform the UI. While our approach is designed to connect metadata with a data discovery system rather than to inform visualization primitives, the declarative specification (or configuration) of software artifacts and tools is common across domains. The specification we propose for Humboldt can be adapted to the business needs. However, some fundamental elements should be included in any implementation of the Humboldt system framework. We discuss these elements of the Humboldt specification in the following.

### 4.1 Metadata Providers

Humboldt is designed to support different metadata providers to fetch data based on the metadata at hand. Data fetching can be done using, e.g., materialized views of a database, lookup tables, SQL statements, or ML models. In short, the implementation of the metadata provider is designed to be independent of the Humboldt specification or the rendered data discovery UI. Instead, what metadata providers need to specify is not how the data is retrieved but what type of data to expect. Therefore the Humboldt specification expects a category and name for the metadata provider, the representation of the returned data (i.e., list, embedding, graph, etc.), any input values the metadata provider requires, and an endpoint for the data discovery system to retrieve that data from. In many cases, it is also helpful to provide information about the visibility of the metadata provider in different parts of the UI so that the data discovery system does not get overloaded. In turn, in our implementation of Humboldt, metadata providers are specified as shown in Figure 3.

The number of metadata providers might quickly grow beyond the point where simultaneously exposing all these providers to the user is feasible. To mitigate this, we enable the specification of a metadata provider type to group metadata providers. The metadata provider's name helps disambiguate individual items in these groups, whereas the description provides more accurate information on the functionality of the provider. Metadata comes in different types and forms, such as annotations (e.g., ownership information, data type), interaction data (e.g., view count, creation date), or relatedness (e.g., data lineage, similarity, joinability, unionability). Hence, this data might also be represented in different ways, e.g., as hierarchical data, where information about child elements is provided, as graph data, where a data artifact's information includes connections, or simply as numeric or textual annotations. To inform the UI about the data representation to expect from the metadata provider, Humboldt's specification includes a representation field. Based on this representation, different views can be generated, as shown in Figure 2. Some metadata providers might rely on user input before fetching data. Therefore, the types of input values and whether that input value is required for the metadata provider to be queried also need to be specified.

## 4.2 Ranking

Once data is retrieved, ranking can help greatly with data discovery. However, ranking weights are unlikely to stay the same throughout the lifespan of a data discovery system. New metadata fields can become available for ranking, and ranking priorities can change. One might, e.g., want to highlight new data artifacts for systems that have just been set up and converge to trusted and frequently used data once a discovery system gets more mature, or even base data ranking on weights obtained from an ML model that is continuously updated. Additionally, metadata providers may have different metadata fields or preferences, which in turn can be useful for ranking. Thus, we add ranking weights to Humboldt's specification to extend the flexibility of Humboldt. While ranking weights can be defined individually for each metadata provider, global ranking weights can be used as a

```
"ranking": [
  {
    "field": "favorite", "weight": 4.3
  },
  {
    "field": "views", "weight": 1.5
  }
]
```

**Listing 1: Humboldt specification enables custom ranking weights to be assigned to metadata fields without changing the underlying ranking algorithm.**

fallback. Whenever multiple metadata providers are combined e.g., for advanced search queries, the ranking results need to be combined. Therefore, we employ a numeric ranking where metadata fields are specified alongside a ranking weight (Listing 1).

Values of metadata fields are multiplied with the ranking factor, which results in an overall ranking score that can be combined between metadata providers. This way, the ranking algorithm code in the UI does not need to be updated whenever ranking weights change or new metadata becomes available. Instead, updating Humboldt's specification is sufficient for updating the ranking algorithm across the data discovery system.

## 4.3 Application-specific Content

Depending on the use case, one might want to add application-specific content to their specification to inform their data discovery UI. The advantage of adding such content in the Humboldt specification over simply hard-coding is that it can refer to other elements in the specification. For example, in our use case (Section 6), we wanted to provide a custom home page for the data discovery system depending on the team memberships of the discovery system's user. This home page should show specific metadata providers for individual teams and, therefore, was defined as shown in Listing 2.

This custom content is flexible and can be tailored to an organization's needs. However, this also means that it is not fully specified and, as such, not transferable. A data discovery UI needs to know about the *fields* and their structure; otherwise, it cannot use them. We think that adding such custom information to the Humboldt specification helps inform specific aspects of a data discovery UI that are not the same for different systems. If a custom field defined in the specification is not supported by the UI implementation, it is ignored.

## 4.4 Customization

A primary advantage of a specification-based interface for data discovery is its configurability. The configuration options that Humboldt's specification provides include the list of metadata providers and their availability in different views. Users might customize this configuration based on their specific needs by modifying the specification directly or through a UI, as shown in Figure 4. Additionally,

```
"custom": [
  "field": "home",
  "content": [
    {
      "data": ["Team", "Favorites", "Shared"],
      "name": "A Team"
    },
    {
      "data": ["Team", "Endorsed", "Recommended"],
      "name": "Research"
    },
  ]
]
```

**Listing 2: Custom content in the Humboldt specification can refer to metadata providers and is more flexible than hard-coded references. However, the UI implementation has to know about the custom fields and their types to display them.**

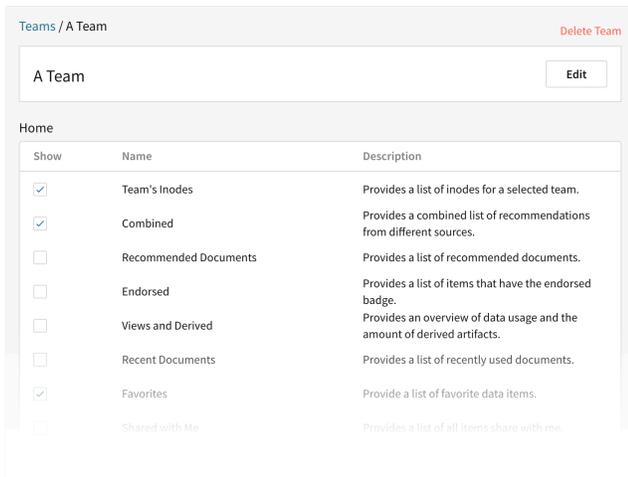

Figure 4: Configuration of the team homepage in a data discovery UI generated by Humboldt. Team administrators can select from the list of metadata providers to enable their visibility and use in the data discovery UI.

developers of metadata providers might add newly implemented metadata providers to Humboldt's specification while removing ones that are not supported anymore. Administrators of an organization that uses a data discovery system can configure which metadata providers they want to use and where these providers are available in the data discovery system. Similarly, individuals using the data discovery system can hide and reorder the metadata providers that they have access to. Finally, with the custom content described in the previous subsection, a team manager within an organization might even configure the recommendations and information surfaced specifically for their team members (Figure 4).

## 5 INTERFACE CONSTRUCTION

A UI for data discovery can be automatically constructed using Humboldt's specification as outlined in Section 4. This includes three main aspects of data discovery: overviews, exploration, and search. We will explain how Humboldt's specification can inform all three aspects to support data discovery.

### 5.1 Overviews

Data overviews are required whenever the user has no clear idea where to start their analysis. In such cases, Humboldt's metadata providers can be used to obtain data overviews. These overviews can each provide different windows into the data warehouse or lake as they are generated from different metadata providers and, thus, based on various metadata fields. For example, the frequency or recency of use, usage logs by similar user types or team members, or information about the content can all be used to surface data to the user. Whenever new metadata providers become available, they can be added to Humboldt's specification, automatically generating views for them. Different users might rely on separate metadata in their day-to-day analysis. With Humboldt, the views that are shown to a user can be configured to match these needs. Since Humboldt's specification includes different data representation types, views can be designed to represent the data at hand best, as shown in Figure 3.

Through a specification of the representation type, views can follow the nature of the data. As such, data overview representations might be visualized e.g., as lists, embeddings, graphs, or hierarchies (Figure 6).

### 5.2 Exploration

Overviews cannot always precisely surface the data that helps the user the most. To find this data, users need to be able to explore from these overviews that serve as entry points to the data discovery system. Humboldt helps with this exploration as it can automatically surface further data upon interaction. Whenever a user interacts with a data element, the metadata of this element can be used to inform and surface more metadata providers. Similar to the Humboldt's overviews, this exploration approach utilizes the specification to inform the data discovery UI about which metadata providers to present. This way, starting from one data artifact, the user can explore related or similar data. For instance, Humboldt might extract ownership, annotation, and usage information from a selected data table. This information can be used to bring up more data from that owner, data with similar annotations, or data with similar usage characteristics. In turn, the user can start their exploration at one data artifact, explore data that is similar with respect to different metadata attributes, and finally land at the data that is most helpful for their analysis.

### 5.3 Search and Filters

Users sometimes have a concrete idea about the data they are interested in or want to filter the data presented to them through the various metadata providers. This is where a query-based search and filtering interface can help. Based on Humboldt's specification, we can generate a query language for a search interface. In Humboldt, search and filters use the same metadata providers that are used to generate views to fetch data. Using this query language, users can utilize all the available metadata providers for search in addition to conventional text-based search. These metadata provider-based search elements are synthesized from Humboldt's specification and support the selection of input elements to inform the metadata providers. As such, If a metadata provider requires an input value, Humboldt can recommend plausible values based on the specified input type as shown in Figure 5. For example, if a user wants to search for documents owned by a specific user, Humboldt looks up that metadata provider and informs the user that they need to enter a user id. Each query element returns a list of data artifacts. Combining multiple query elements in a search query allows for an arithmetic combination of different search queries and their resulting data artifact lists. In our use case, we implemented two logical connectors (*and*, *or*) while supporting bracketed queries and negation for further specification.

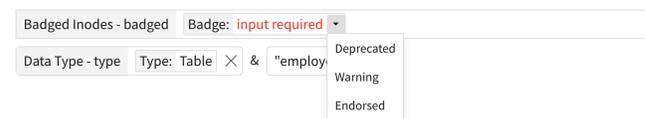

Figure 5: Search queries can be formed by combining free text keywords, metadata field-value pairs, and logical operators. Humboldt uses metadata specifications to determine admissible field-value pairs and compositions. The top bar shows a newly entered query element, while the bottom shows the active search query.

The difference between search and filters is the set of data artifacts it is performed on. Thus, when using the search query while inspecting a view, the displayed data (i.e., from the specific metadata provider) is filtered. In contrast, all data available to the user is considered when searching globally. In the following, we will interchangeably use the terms *search* and *filter*.

We implemented two different search interfaces to show the flexibility that Humboldt provides for implementing UI elements and interaction concepts. These search interfaces vary in the way search queries are presented and entered. We implemented a pill-based query representation (Figure 5) and a prefix-based textual query language (e.g., *":recent_documents() & bit"*, which is a combination of a metadata provider and text query).

## 6 USE CASE

To show how our Humboldt can generate a rich data discovery UI, we used it as a framework to create one in Sigma Workbook.

### 6.1 Metadata Providers

To show the flexibility of Humboldt with respect to different metadata providers, we added a large set of available metadata providers to the Humboldt specification for this use case. Some of the providers we integrated are depicted in Figure 2.

The metadata associated with the providers became available as new views for discovery and in the search interface. Depending on the input values required for the metadata provider, Humboldt automatically determines whether the metadata provider has all the information needed for fetching data. The data returned by the metadata provider is visualized using the representation specified in the Humboldt specification as shown in Figure 6.

### 6.2 Discovery Views

While the data discovery UI can be automatically generated from Humboldt's specification, there must be visual primitives for the different data representations that the specified metadata providers return. In this use case, we implemented six different visual representations for data. As such, data can be represented via tiles, in a list, hierarchically, in a graph, as categories, and in an embedding view (Figure 6). We briefly describe these views below.

**Tiles.** The tiles view displays data as boxes (tiles) in a grid. Tiles can be ordered via specified ranking weights and provide an overview of available data while not overwhelming the user with many small data artifacts.

**List.** To display a large number of data artifacts, the tile representation may not be ideal as it requires a sizeable screen space. We implemented an ordered list as another visual representation of data artifacts. This list can be ordered based on the specified ranking or by clicking any columns in the list view.

**Hierarchy.** The hierarchy (tree) view enables the navigation of one-to-many relationships defined by metadata. For example, it is often helpful to have metadata specifying which other data artifacts use a given data artifact (e.g., a table can be used to create a visualization, which in turn can be embedded in a dashboard, where the table, visualization, and dashboard are examples of data artifacts.) While this view currently uses tiles to represent the nodes at each level, it is easy to add list-based or hybrid variants, dynamically switching between the most effective node rendering depending on viewing constraints. The hierarchy view supports traversing hierarchies of arbitrary depths.

**Graph.** The graph view supports displaying graph-structured metadata (e.g., join paths) that describes how data artifacts are related to each other. In addition to information about data artifacts (i.e., nodes), the graph view expects the metadata to contain information about how they are connected.

**Categories.** Perhaps the most common metadata is categorical metadata, which describes a category of data artifacts. The category can be, for example, a label, a data type, or an owner name. The categories view enables an effective exploration of data artifacts based on their categories while providing an overview of the available categories.

**Embedding.** The embedding view supports the display of data artifacts using their two-dimensional positional encoding metadata. Projection (embedding) views are generally useful for exploring large numbers of high-dimensional data points. The embedding view shows data artifacts on a two-dimensional canvas as circles and, therefore, expects the x and y coordinates to be included in the data artifacts metadata. We anticipate embedding metadata to be increasingly common with AI applications for data management and analysis, where learned representations of data artifacts and their elements are typically computed and used.

### 6.3 Interactive Exploration

Humboldt-generated UIs support interactive exploratory discovery that enables users to navigate data using the metadata. Consider a data artifact selected by a user as shown Figure 7; based on the metadata associated with the data artifact, the user can view the other data artifacts with the same badges (cf. Figure 2, Badged), the same owners (cf. Figure 2, Owned By), and the same data type (Figure 2, Type).

### 6.4 Search

If a data system already supports text-based search, Humboldt makes it easy to integrate prefix-based search. As shown in Figure 5, complex search queries can be constructed from text-based and metadata provider-based search query elements. This combination is realized through logical operators. The search functionality obtains its data from the same metadata providers as the discovery views that are specified in Humboldt. It determines the admissible values for metadata constraints in the query based on the Humboldtspecification and helps users dynamically select from these values while entering the query (Figure 5). Whenever a search query is entered, results are shown in a new search tab (Figure 7, B) using the list view. The keyword-based search can also be used to filter exploratory views. For example, users can use the search functionality to show only the matching data artifacts in the joinability graph (Figure 2, Joinable).

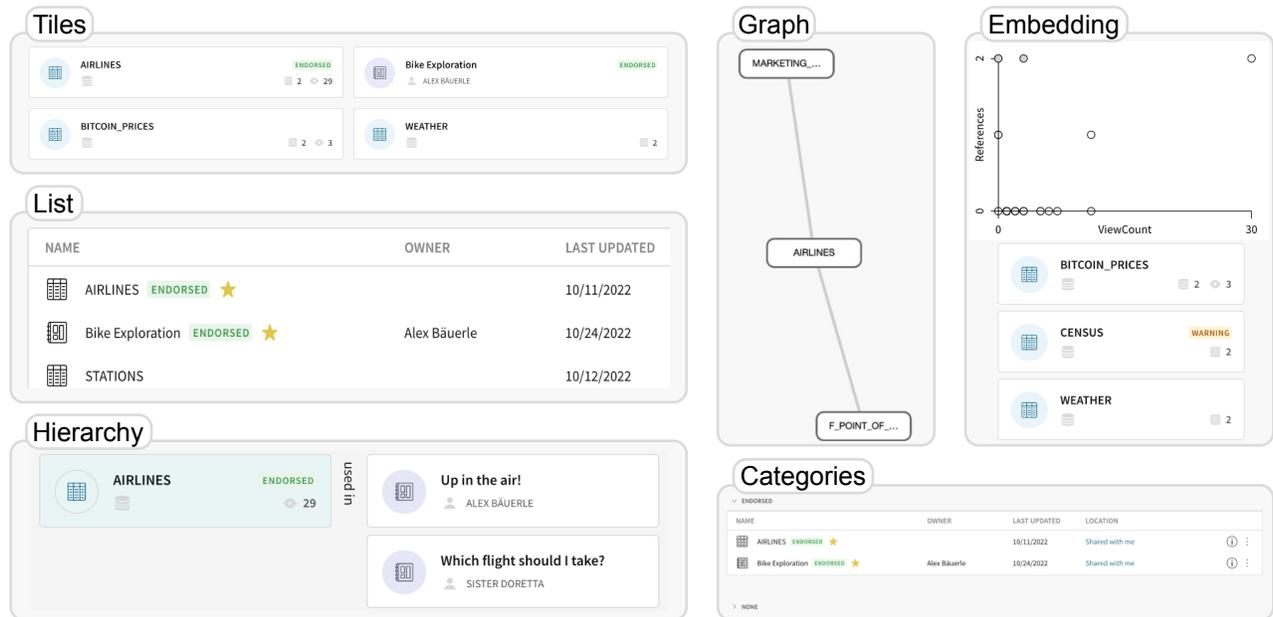

Figure 6: Examples of data discovery interfaces dynamically generated by Humboldt. Depending on metadata provider specification, data can be displayed as tiles, a list, a hierarchy, a graph, an embedding (scatter plot), or categories.

## 7 EVALUATION

As our use case presented in Section 6 shows, the Humboldt framework enables the integration of various metadata providers into a data discovery UI. To evaluate how end users are affected by a data discovery system implemented based on Humboldt, we recruited six users at Sigma Computing for a first-use study. Our participants were recruited from the Sales Engineering department to ensure familiarity with customer needs as well as the system. None of our participants had seen the Humboldt UI before this evaluation.

### 7.1 Setup, Tasks, and Protocol

Participants were first given a short overview of the data discovery UI we introduced in Section 6. Then, they were asked to freely explore the UI for a few minutes to familiarize themselves.

We designed Humboldt to support expressivity, composability, and (re)configurability. We asked participants to perform four tasks that were devised to evaluate the UI in these three aspects. To assess how well the data discovery UI achieves expressivity and composability, three of the tasks required participants to use the different views as entry points into data space, navigate starting from a specific data artifact to perform an exploratory search, and utilize search and filtering functionality to find a data artifact of interest in directed search, respectively. While testing these aspects of the data discovery UI, participants assumed the role of a typical user of the system. After these three tasks, we asked participants to switch to a team admin role to test Humboldt's configurability. The fourth task was configuring the data discovery UI for a team they were managing.

Before each task, the experimenter verbally explained the task and answered questions from participants. We used a think-aloud protocol during task completion, whereas the examiner transcribed notes. After completing the four tasks, participants completed a questionnaire eliciting feedback on the UI's affordances. Each session lasted about 30 minutes.

**Task 1.** The first task was to *"find table AIRLINES, which has the endorsed tag."* It was aimed to evaluate the effectiveness of the metadata-based overviews. Participants could use the different entry points to find a specific table based on a metadata field. Hence, they had to understand the different metadata-based representations before finding a particular datapoint .

**Task 2.** For the second task, we wanted to evaluate the effectiveness of the Humboldt generated UI's exploratory data discovery support. The task was, therefore, to *"find other elements that are similar to the table w.r.t. type or badge."* Participants could use added views to find data related to a selected data artifact.

**Task 3.** The third task aimed to assess the usability of the search and filtering functionalities and the ability to compose different metadata conditions using the keyword query interface. Therefore, we asked participants to *"find all workbooks created by user John Doe"* from the view they were currently on. To do this, participants needed to use search and filtering and enter their query using metadata-based filters.

**Task 4.** Finally, we asked participants to take up the role of a team administrator and change the discovery views shown on a team's home page: *"assume you are the administrator of A Team in your organization and set the team's home page to your preferred content."* This task required them to reconfigure Humboldt based on their team's needs.

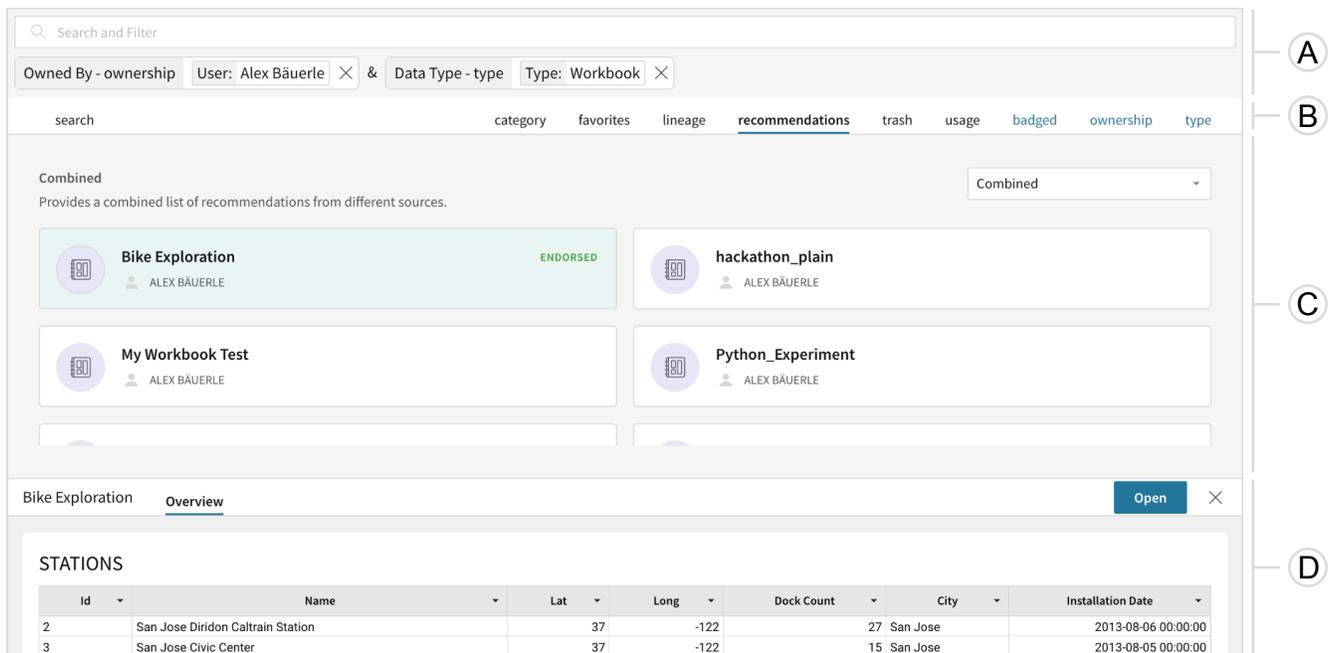

Figure 7: Screenshot of the data discovery UI built in Sigma Workbooks using Humboldt. Humboldt can automatically generate data discovery UIs supporting directed and exploratory searches. (A) The keyword query search interface allows the use of metadata fields and values together with free text keywords. (B) Overviews based on the available metadata are organized into tabs. (C) The active tab displays the data provided by the selected metadata category (i.e., recommendations). The data shown here is filtered using the query entered in the keyword query search field. (D) A content preview is shown when an individual data artifact is selected. In this case, the data artifact is a table, and the preview shows a snippet of the table.

### 7.2 Results

All participants were able to complete all four tasks. They completed Task 1 without help while following different routes. Three participants jump-started with the keyword search and later discovered the metadata-based views to complete the task. The rest directly started from data discovery views to carry out the task. For Task 2, we reminded three participants that new data discovery views might be populated on selecting a data artifact, depending on the metadata. When performing Task 3, half of the participants missed the first condition and did not filter out only workbooks. After reminding them of that, they were able to complete the task. Finally, for Task 4, two participants needed help finding the team configuration setting but had no problem configuring a team's page.

We asked our participants to provide feedback and describe their actions while performing the tasks. After completing all tasks, we also asked them about the UI components they interacted with, focusing on the search functionality, the discovery views, and Humboldt's customizability.

**Views.** Participants expressed they *"like the different views a lot"* (P6) and that *"it makes sense to have multiple views to present data"* (P5). *"I like the scatter plot for finding outliers in usership"* (P6), *"I like the usage graph"* (P4), and *"I would want more of the categorical views"* (P2). One participant indicated they *"would probably mainly use search as I don't want to dig and find stuff"* (P4), and another would *"want a global overview that I can filter"* (P5). For such users, the UI can be configured to show just a small number of suggestions and direct everything else to search.

The feedback suggests that data discovery is user—and task-dependent; different users have different preferences and needs shaped by their experience and roles. Having data discovery facets based on metadata providers and supporting different search paradigms is desirable for users, which is facilitated by the expressivity of Humboldt in this case.

**Exploration.** Participants also *"like the ability to link selected things to new metadata providers"* (P1). When exploring, they found *"the preview of selected data really cool"* (P3), from where they could explore related data artifacts. They also gave positive feedback about the ability to compose views with filters to explore further within a metadata provider-driven views. For example, *"for large amounts of data, the usage view combined with filters would be constructive"* (P3).

**Search and Filtering.** Participants heavily used the search query interface that Humboldt builds based on its metadata providers. They commented that the *"complex search queries I can construct are very useful"* (P6), *"especially for more technical users"* (P1). Humboldt generates the query language based on the specification of metadata providers and provides autocomplete suggestions for admissible prefixes and values as the user types the query. Participants liked that *"you don't have to learn the syntax to be able to search"* (P2) and *"the search can suggest potential inputs"* (P1). Two participants also mentioned that they would prefer *"a freeform search instead*

*of the query language"* (P3). At the same time, three other participants argued against freeform search, commenting that *"people like freeform search in theory, but in practice, it can be frustrating"* (P6). We concur that freeform (natural language) search has to work well to be useful, for which the current autoregressive language models provide an opportunity (e.g., [22, 24]). Nonetheless, natural language is inherently ambiguous and is generally good for high recall results but not necessarily efficient for high precision. A compromise could be to use our query language that is generated from Humboldt's specification and let users *"convert the search into a free text formula"* (P4) or even a freeform search. The participants' comments above further highlight users' varying data discovery needs. We postulate that experienced users prefer high-precision interfaces like query-based search, while others benefit from high-recall data discovery functionalities at first.

**Customizability.** Participants reported that Humboldt's customizability *"can help find an entry into otherwise overwhelming amounts of data"* (P1) that is tailored to a user's specific analysis needs, and *"can be a way to start the analysis process much more efficiently"* (P6). Participants thought *"configurability would probably help customers"* (P5), as business users are used to customizing their workflow, e.g., *"in GSuite I also customize how my calendar looks"* (P5). While participants only experimented with customizing a team's landing page, they *"would imagine a whole section for customizing the interface"* (P2). This included configuring the metadata providers that are shown as views over the metadata providers used in the search query language and the order in which metadata providers are shown. Participants also mentioned that they either *"would not want to touch the configuration"* (P4), or *"would also want to customize how things look (e.g., list vs. tiles)"* (P5), and *"would like to customize the order in addition to what is shown"* (P6). Overall, they found that *"customizability would definitely be useful"* (P6), underlining the value of extending or adapting the data discovery UI to specific needs.

**Areas of Improvement.** While the feedback for Humboldt was positive overall, participants also offered ideas for improvement. For example, they *"sometimes do not know what a metadata provider means"* (P4), and would like *"a more detailed description of individual metadata providers"* (P1). Views that required an input could benefit from *"an indication of the current input to a metadata provider"* (P4). They also commented on further enriching the current exploratory search support, e.g., through *"clicking on an owner to see their data artifacts"* (P5). Participants also suggested improving the layout of some UI elements: *"it would be more intuitive if related data were shown closer to selected metadata provider-based views"* (P1), *"it was not clear for me that some metadata providers were based on selected content"* (P2), and *"customization is a bit buried right now"* (P2). This paper focused on the Humboldt system framework functionalities rather than the user experience design of the generated data discovery UIs. The feedback emphasizes that they both go hand in hand and that more work on the design with iterative feedback cycles can further improve the user experience with data discovery.

**Post-Study Questionnaire.** After the main study, we asked our participants to complete a short survey about the discovery UI generated by Humboldt. They rated 12 statements (Figure 8) from 4 categories regarding the UI on a 5-point Likert scale that ranged from 1 (*Strongly Disagree*) to 5 (*Strongly Agree*). Their ratings were positive about all aspects of the UI across categories on average (mean:3.97, std:0.85). The participants were the most affirmative about the metadata support for search (mean:4.33, std:0.75) and previews (mean:4.33, std:1.11). On the other hand, they judged finding data views (mean:3.33, std:0.75) and layout design (mean:3.50, std:0.96) the least affirmatively. These suggest Humboldt generated discovery UI can benefit from improved design for ease-of-use and comprehension beyond its initial focus on data discovery affordances. Also, all but one participant rated Humboldt's configurability support enabling customization (mean:4.17, std:0.69) and extension (mean:4.17, std:0.69) helpful.

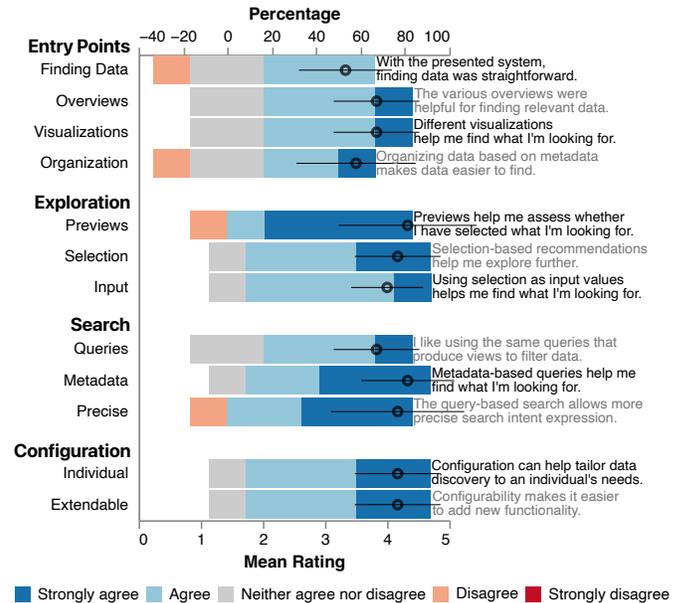

Figure 8: We elicited feedback from 6 participants responding to 12 questions in 4 categories. The top axis and bars correspond to the percentage of positive/negative answers, while the bottom axis and circles show the mean and std for the Likert scale ratings. Overall, the feedback was positive across all categories, suggesting additional evidence that Humboldt achieves its design goals. Entry points received the most mixed responses, primarily due to the layout of UI elements, which can be easily improved as Humboldt enables quick experimentation with data discovery UIs.

## 8 DISCUSSION

Results of our evaluation validate the design goals of Humboldt and support their overall successful execution. Feedback from the participants suggests different users and teams have different data discovery needs. Therefore, the ability to support multiple data discovery paradigms, compose, customize, and extend the data discovery functionalities while leveraging metadata is crucial. Metadata is how organizations augment data with their business considerations and domain know-how.

We conducted the user study to elicit in-context feedback on Humboldt replicating its usage in practice. The discovery UI generated as part of Sigma Workbook is fully functional and integrated, enhancing the ecological validity of our study and its results. As in any user study, the number and the pool of participants who took part in our user study constitute a biased sample of all user profiles and goals relevant to data discovery UIs in data systems. Nonetheless, we believe our participants' regular interaction with users from diverse domains to improve data-centric tasks and their understanding of data discovery needs around these tasks based on these interactions enabled gathering helpful feedback and insights.

Usability remains a significant challenge in user-facing data systems [8]. Natural language interfaces enabled by LLMs also provide a potential [22, 24] to improve data discovery interfaces. Future work can explore combining the precision of query-based search enabling metadata constraints with the high recall of natural language. Finally, we believe that Humboldt could be useful beyond BI systems. Future work can also explore its integration with other interactive data systems for better data discovery.

## 9 CONCLUSION

Metadata provides valuable business and usage context for data. Therefore, supporting metadata as a first-class citizen in data discovery interfaces is a good idea, enabling users to search and navigate data using metadata attribute values. Organizations, teams, and individual users have different data discovery needs based on their domains, tasks, and experiences. It is also a good idea to endow data systems with the ability to easily integrate different metadata sources for data discovery, support multiple data discovery paradigms and views, and reconfigure and extend for custom and domain-specific data discovery needs. The Humboldt framework we introduced here operationalizes these ideas.